\documentclass[journal]{IEEEtran}

\usepackage{cite}
\usepackage{amsmath,amssymb,amsfonts,amsthm}
\usepackage{graphicx}
\usepackage{textcomp}
\usepackage{algorithm}
\usepackage{algpseudocode}
\usepackage{cancel}
\usepackage{hyperref}
\usepackage{cleveref}
\usepackage{graphicx,epsf}
\usepackage{enumerate,color}
\usepackage[mathcal]{euscript}
\usepackage[dvipsnames]{xcolor}
\usepackage{tikz}
\usetikzlibrary{spy}

\newcommand{\R}{\mathbb{R}}

\newcommand*\Let[2]{\State #1 $\gets$ #2}
\DeclareMathOperator*{\argmin}{arg\,min}
\newtheorem{theorem}{Theorem}
\newtheorem{corollary}{Corollary}

\newtheorem{proposition}{Proposition}
\begin{document}

\title{Model-corrected learned primal-dual models for \\ fast limited-view photoacoustic tomography}

\author{Andreas Hauptmann~\IEEEmembership{Senior Member,~IEEE}, and Jenni Poimala% <-this % stops a space
\thanks{This work has been supported by the Academy of Finland projects 338408, 346574, 353093. The authors would like to thank the Isaac Newton Institute for Mathematical Sciences, Cambridge, for support and hospitality during the programme \emph{Rich and Nonlinear Tomography} where work on this paper was undertaken, supported by EPSRC grant no EP/R014604/1.}
\thanks{JP and AH are with the Research Unit of Mathematical Sciences, University of Oulu, Finland.}
\thanks{AH is also with the Department of Computer Science, University College London, UK.}
}

% note the % following the last \IEEEmembership and also \thanks - 
% these prevent an unwanted space from occurring between the last author name
% and the end of the author line. i.e., if you had this:
% 
% \author{....lastname \thanks{...} \thanks{...} }
%                     ^------------^------------^----Do not want these spaces!
%
% a space would be appended to the last name and could cause every name on that
% line to be shifted left slightly. This is one of those "LaTeX things". For
% instance, "\textbf{A} \textbf{B}" will typeset as "A B" not "AB". To get
% "AB" then you have to do: "\textbf{A}\textbf{B}"
% \thanks is no different in this regard, so shield the last } of each \thanks
% that ends a line with a % and do not let a space in before the next \thanks.
% Spaces after \IEEEmembership other than the last one are OK (and needed) as
% you are supposed to have spaces between the names. For what it is worth,
% this is a minor point as most people would not even notice if the said evil
% space somehow managed to creep in.

% The paper headers
\markboth{Hauptmann \& Poimala, Model-corrected learned primal-dual models, April 2023}%
{Shell \MakeLowercase{\textit{et al.}}: Bare Demo of IEEEtran.cls for IEEE Journals}
% The only time the second header will appear is for the odd numbered pages
% after the title page when using the twoside option.
% 
% *** Note that you probably will NOT want to include the author's ***
% *** name in the headers of peer review papers.                   ***
% You can use \ifCLASSOPTIONpeerreview for conditional compilation here if
% you desire.

% If you want to put a publisher's ID mark on the page you can do it like
% this:
%\IEEEpubid{0000--0000/00\$00.00~\copyright~2015 IEEE}
% Remember, if you use this you must call \IEEEpubidadjcol in the second
% column for its text to clear the IEEEpubid mark.

% use for special paper notices
%\IEEEspecialpapernotice{(Invited Paper)}

% make the title area
\maketitle

% As a general rule, do not put math, special symbols or citations
% in the abstract or keywords.
\begin{abstract}
Learned iterative reconstructions hold great promise to accelerate tomographic imaging with empirical robustness to model perturbations. Nevertheless, an adoption for photoacoustic tomography is hindered by the need to repeatedly evaluate the computational expensive forward model. Computational feasibility can be obtained by the use of fast approximate models, but a need to compensate model errors arises. In this work we advance the methodological and theoretical basis for model corrections in learned image reconstructions by embedding the model correction in a learned primal-dual framework. Here, the model correction is jointly learned in data space coupled with a learned updating operator in image space within an unrolled end-to-end learned iterative reconstruction approach. The proposed formulation allows an extension to a primal-dual deep equilibrium model providing fixed-point convergence as well as reduced memory requirements for training. We provide theoretical and empirical insights into the proposed models with numerical validation in a realistic 2D limited-view setting. The model-corrected learned primal-dual methods show excellent reconstruction quality with fast inference times and thus providing a methodological basis for real-time capable and scalable iterative reconstructions in photoacoustic tomography.
\end{abstract}

% Note that keywords are not normally used for peerreview papers.

% For peer review papers, you can put extra information on the cover
% page as needed:
% \ifCLASSOPTIONpeerreview
% \begin{center} \bfseries EDICS Category: 3-BBND \end{center}
% \fi
%
% For peerreview papers, this IEEEtran command inserts a page break and
% creates the second title. It will be ignored for other modes.
\IEEEpeerreviewmaketitle

\section{Introduction}
\IEEEPARstart{L}{earned} iterative reconstructions are highly popular due to their capabilities to combine model information and learned components to achieve state-of-the-art results with robustness to model perturbations \cite{arridge2019solving,monga2021algorithm,mukherjee2023learned,kamilov2023plug}. In particular, so-called end-to-end approaches that unroll an iterative scheme for a fixed number of iterations and train all iterates jointly provide superior reconstruction quality in many applications. 
Nevertheless, applicability of such end-to-end approaches is largely limited to two-dimensional scenarios with small image sizes where training and evaluation is feasible. For high resolution imaging, or even three-dimensional data, we encounter several key difficulties. First of all, memory demands increase linearly with number of unrolled iterates and secondly, computationally expensive evaluations of the model equations can severely limit reasonable training times. This is especially so in photoacoustic tomography, where the forward model is given by an acoustic wave equation. Additionally, many photoacoustic measurement configurations employ a limited-view geometry making the inverse problem severely ill-posed, necessitating advanced reconstruction approaches and strong priors in the reconstruction process. 

One such class of suitable advanced reconstruction methods consists of primal-dual approaches that have been very successful in classic model-based imaging scenarios \cite{Chambolle2011,sidky2012convex} and in particular its learned version still provides state-of-the-art results \cite{Adler2018}. The combination of a learned update in image (primal) and data (dual) space is particularly powerful, but also necessitates a joint training in an end-to-end fashion leading back to the main limitations of adaption to high dimensions. In particular, the memory demand is even further increased due to the inclusion of an updating network in the measurement space. 

There have been several recent advances to overcome these computational limitations in different settings, roughly following two paradigms: 
\begin{itemize}
    \item[(i)] Uncoupling of network training from the operator evaluation. 
    \item[(ii)] Methodological improvements in the learned iterative approach to negate impact of high computational demand.
\end{itemize} 

In the first category are so-called plug-and-play approaches, where a denoiser is learned offline and subsequently used in an iterative scheme such as ADMM or proximal gradient descent, replacing the proximal operator by the learned denoiser \cite{venkatakrishnan2013plug,zhang2021plug,ryu2019plug,laumont2022bayesian,kamilov2023plug}. 
%The nature of the denoiser makes this approach difficult to apply to limited-view settings. Nevertheless, recent advances using normalising flows(?) enable an application to limited angle X-ray tomography [REF].
Another alternative is a greedy training that requires iterative wise optimality instead of training the unrolled iterates end-to-end \cite{Hauptmann2018,aghabiglou2022deep}, enabling an application to 3D limited-view photoacoustic tomography \cite{Hauptmann2018}. Unfortunately, the joint nature of learned-primal dual methods prohibits to adapt a greedy training strategy. 

The second category is more varied, as approaches may tackle only a specific subproblem. For instance, an approximate model has been used in \cite{hauptmann2018MLMIR,lunz2021learned,smyl2021learning} to speed-up training and deployment of the models. Deep equilibrium (DEQ) models \cite{bai2019deep,gilton2021deep}, aim to reduce memory consumption by reformulating the unrolled algorithm to a fixed-point iteration and differentiating only with respect to the fixed-point equation, allowing for constant memory consumption independent of the amount of iterates. Another work \cite{hauptmann2020multi} aimed to reduce both memory consumption and operator evaluations by formulating the unrolled algorithm as a multiscale approach, reducing the computational cost of forward operator and memory consumption on the lower scales. Finally, invertible neural networks can be used \cite{putzky2019rim,rudzusika2021invertible,orozco2022memory}, eliminating the need to store intermediate states for backpropagation, but the need to evaluate the forward operator remains in an unrolled algorithm.

In addition to the above computational advances, there are recently increased efforts to provide theoretical results for learned image reconstructions \cite{mukherjee2023learned,ryu2019plug,pesquet2021learning,li2020nett}, for instance in the form of convergence results for unrolling approaches. One such convergence guarantee is given for DEQ models to a fixed point by requiring the updating operator to be a contraction \cite{bai2019deep,gilton2021deep}. Stronger notions of convergence can be satisfied, but generally require more restrictive constraints on the learned components, often reducing expressivity and performance of the networks. We refer to \cite{mukherjee2023learned} for an overview of provably convergent approaches for learned image reconstruction.

Motivated by the above developments, this work aims to formulate a model-corrected learned primal-dual model, that can be trained end-to-end for high resolutions using fast approximate forward operators. Additionally, we provide theoretical insights and conditions on fixed-point convergence in a deep equilibrium formulation. 
To achieve this, we start with the classic model-based primal-dual hybrid gradient \cite{Chambolle2011} and replace the expensive forward and adjoint operators with a fast, but approximate version \cite{hauptmann2018MLMIR}. The introduced errors are corrected with an operator correction in dual space, motivated by \cite{lunz2021learned}, and a learned updating operator in primal space. The resulting algorithm is called a model-corrected learned primal-dual (MC-PD). We then proceed to formulate the model-corrected learned primal-dual as a deep equilibrium model (PD-DEQ) and analyze conditions on parameters and networks to provide a contraction on the primal variable. Our main results state that fixed-point convergence can be achieved under a sufficient decrease assumption on the dual variable. 

The developed methods are evaluated extensively for a 2D limited-view scenario. We compare our methods to a post-processing approach with U-Net and a version of the proposed MC-PD that does not use weight-sharing and hence is close to the original learned primal-dual. We evaluate scalability in terms of memory consumption and evaluation times to larger image sizes. Our results show that MC-PD provides the same excellent performance with weight-sharing as well as without. The DEQ models do not perform as well as the MC-PD with a drop in PSNR and SSIM, but providing good qualitative results. Additionally, we discuss challenges when training the DEQ models in the considered limited-view setting and computationally verify the contraction of the updating network. 
Finally, we will provide codes and implementation of our models.

This paper is structured as follows. In sec. \ref{sec:PAT} we introduce the main concepts in photoacoustic tomography and the variational formulation to reconstruction. We then review learned reconstruction approaches and introduce the fast approximate operators used in this study. In sec. \ref{sec:MC-PD} we introduce the model-corrected learned primal-dual and its deep equilibrium formulation. We provide some theoretical insights to proposed models and then proceed to derive conditions on the contraction property. Sec. \ref{sec:CompMethods} discusses computational aspects and data simulation. We present the results in sec. \ref{sec:resultsDisc} and discuss performance and challenges of the proposed models. Finally, sec. \ref{sec:conclusions} provides final conclusions.

\section{Photoacoustic tomography and prior work}\label{sec:PAT}
In photoacoustic tomography (PAT) a short pulse of near-infrared light is absorbed by chromophores in biological tissue \cite{Beard2011,cox2012quantitative}. For a sufficiently short pulse, this will result in a spatially-varying pressure increase $x$ inside the tissue, which initiates an ultrasound (US) pulse (\textit{photoacoustic effect}), that propagates to the tissue surface. 
In this work we only consider the latter acoustic problem of photoacoustic tomography. That is, the forward model of acoustic propagation is modeled by an initial value problem for the acoustic wave equation \cite{kuchment2011mathematics,wang2015photoacoustic}
\begin{equation}
\label{eqn:PATfwd}
(\partial_{tt} - c^2 \Delta) p(\mathbf{x},t) = 0, \,\, p(\mathbf{x},t_0) = x(\mathbf{x}), \,\, \partial_t p(\mathbf{x},t_0) = 0, 
\end{equation}
with $\mathbf{x}\in\R^2$ and $t_0=0$ denotes the initial time.
We measure the pressure field on the boundary of the computational domain $\Omega$ for a finite time window. The measured data can then be modeled by a linear operator $\mathcal{M}$ acting on the pressure field $p(\mathbf{x},t)$:
\begin{equation}
y = \mathcal{M} \, p_{|\partial \Omega \times (0,T)}. \label{eqn:Measurement}
\end{equation}
The acoustic propagation \eqref{eqn:PATfwd} and the measurement operator \eqref{eqn:Measurement} define the linear forward model
\begin{equation}\label{eqn:invProb}
Ax=y
\end{equation}
from initial pressure $x\in X$ to the measured time series $y\in Y$. This represents the ideal accurate forward model and can be simulated, e.g., by a pseudo-spectral time-stepping model as outlined in \cite{treeby2010,Treeby2012}.

\begin{figure*}
    \centering
    \includegraphics[width=0.9\linewidth]{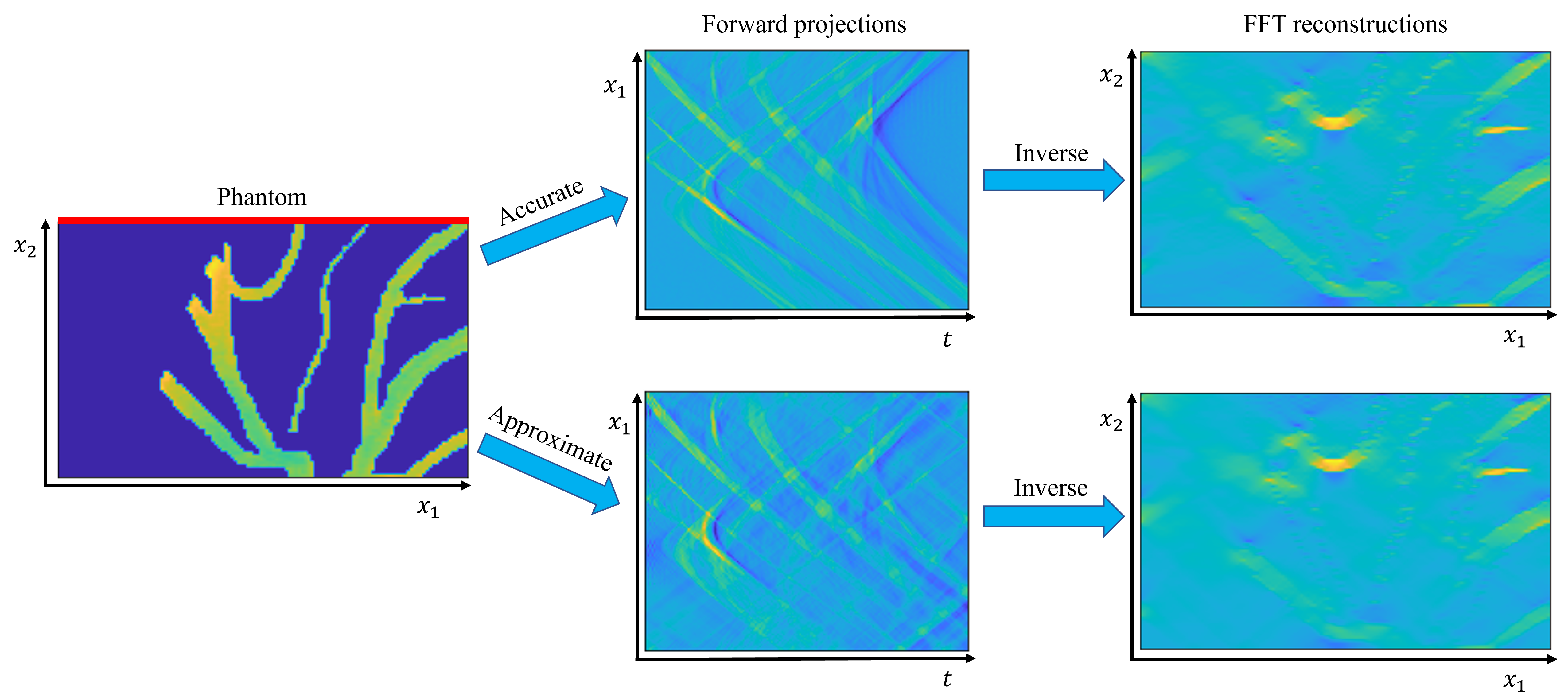}
    \caption{Measurement setup and illustration of the fast models. The phantom (left) left is measured from a one-sided sensor on the top (red line). The ideal measurement is shown in the middle top. The fast approximate forward map is shown in the bottom and contains clear aliasing artifacts. Applying the inverse mapping provides reconstructions with strong limited-view artifacts. One can see on the bottom right, that the obtained reconstruction from the approximate model is smoother compared to the top.  }
    \label{fig:dataGeneration}
\end{figure*}

If the measured data is sparse or the detection geometry is limited to only a part of the boundary the inverse problem becomes severely ill-posed \cite{paltauf2007experimental}, such as in the limited-view scenario considered here, see Figure \ref{fig:dataGeneration} for an illustration of the measurement setup.
This necessitates regularization and advanced priors to compensate for the lost information, which can be formulated in a variational setting \cite{arridge2016accelerated,sandbichler2015novel,boink2018framework}. That is, given measured data $y$ we obtain a reconstruction as the minimizer of
\begin{equation}\label{eqn:variational}
J(x) = \|Ax-y\|_2^2 + \lambda R(x),
\end{equation}
where the first term measures the data-fit and the second term regularizes the problem and incorporates \emph{a priori} knowledge one might have about the target, $\lambda>0$ balances the influence of both terms. Solutions of \cref{eqn:variational} can then be computed iteratively by a suitable optimization algorithm. For instance, using a proximal gradient scheme that allows for non-differentiable $R(x)$: given an initial reconstruction $x_0$ one can iteratively minimize \cref{eqn:variational} by the updating equation
\begin{equation}\label{eqn:proxgrad}
x_{k+1}=\text{prox}_{R,\lambda\gamma_k}\left(x_k - \gamma_k A^*(Ax_k-y)\right),
\end{equation}
with suitable step-length $\gamma_k>0$. The proximal operator projects iterates to solutions admissible by the regularisation term $R$, by solving a corresponding denoising problem
\begin{equation}\label{eqn:proxOp}
\text{prox}_{R,\alpha}(x) = \argmin_{u} \left\lbrace R(u) + \frac{1}{2 \alpha} \|u - x\|_2^2 \right\rbrace. 
\end{equation}
Alternatively, utilizing the dual space we can formulate a min-max optimization problem in primal and dual space. This leads to the highly successful primal-dual hybrid gradient (PDHG) method, \cite{Chambolle2011}. Here, the update equations for the primal variable $x\in X$ and dual variable $q\in Y$ to solve \eqref{eqn:variational} are
\begin{align}
    q_{k+1} &=\frac{q_k + \sigma (A \widetilde{x}_k - y)}{1+\sigma} \label{eqn:PDHG-dual}, \\
    x_{k+1} &=\text{prox}_{R,\lambda\tau}\left(x_k - \tau A^* q_{k+1}\right), \label{eqn:PDHG-primal} \\
    \widetilde{x}_{k+1}&= x_{k+1} + \alpha(x_{k+1} - x_k), \label{eqn:PDHG-relaxation}
\end{align}
where $\sigma,\tau>0$ and $\theta\in[0,1]$. Given the operator norm of the forward operator $L:=\|A\|=\|A\|_{X\to Y}$ convergence to a saddle point for $(q,x)$ is shown when $\sigma\tau L^2<1$ and $\theta=1$ \cite{Chambolle2011}.
Let us point out, that \cref{eqn:PDHG-dual} corresponds to the proximal operator for the dual space, i.e., the least-squares data fidelity term in \cref{eqn:variational}. 
In the following we will base our method on the formulation in eqns. \eqref{eqn:PDHG-dual} and \eqref{eqn:PDHG-primal}.

\subsection{Learned model-based approaches to reconstruction}
Learned reconstruction methods are increasingly popular in medical imaging, as they offer the possibility to combine strong data driven prior information with hand-crafted model components \cite{arridge2019solving}, while also providing a considerable speed-up of reconstruction times. 

Most successful in terms of image quality and robustness are so-called model-based techniques that combine data-driven components with model components. That means, forward and adjoint operator are used within the network architectures \cite{Adler2017,Adler2018,hammernik2018learning,Hauptmann2018}. A prominent class are unrolled iterative methods that follow \cref{eqn:proxgrad} and, e.g., reformulate the updating equations to
\begin{equation}\label{eqn:learnedProx}
x_{k+1}=G_{\theta_k}\left(x_k - \lambda_k A^*(Ax_k-y)\right).
\end{equation}
That is, the proximal operator is replaced by a learned updating network $G_{\theta_k}$ with parameters $\theta$, where the network weights are learned from supervised training data. Note, that in the above formulation the network weights are allowed to change in each iteration. We can also use weight-sharing and fix the parameters $\theta$ globally for all updates. Having a fixed updating operator allows for a more accessible analysis of the iterative reconstruction by constraining the network (weights), or the architecture of the updating operator \cite{mukherjee2023learned}. 

A related approach are so-called plug-and-play methods \cite{kamilov2023plug} that use a separately learned denoiser $G_\theta$, which is subsequently used in the iterative process. Note, also here the denoiser is fixed for all iterates. 
The nature of the network $G_\theta$  to act as a denoiser primarily limits this approach to inverse problems of denoising type, in contrast to the limited-view problem we encounter here. Nevertheless, recent advances show that if $G_\theta$ is given by a generative network one can obtain high-quality reconstructions and uncertainty maps in a related sampling framework \cite{liu2022dolce}.

Similarly to the classic model-based reconstruction in eqs. \eqref{eqn:proxgrad}-\eqref{eqn:PDHG-primal} we can extend the learned iterative approaches into a learned primal-dual method \cite{Adler2018} by replacing the proximal operators in primal and dual space with learned operators. The learned updates are then given in a general form as 
\[
\begin{split}
    q_{k+1} &=F_{\phi_k}\left(q_k, A x_k, y\right) \\
    x_{k+1} &=G_{\theta_k}\left(x_k, A^* q_{k+1}\right).
    \end{split}
\]
The learned primal-dual scheme has shown superior performance for X-ray CT, but training is necessary to be performed in an end-to-end manner for all iterates jointly, due to the coupling of primal and dual updates, which can pose a major computational challenge. Some advances have been made by using invertible networks \cite{bajic20223d,moriakov2022lire} to reduce memory costs.

Nevertheless, in photoacoustic tomography the application of learned model-based techniques has been very limited \cite{Hauptmann2018,hauptmann2018MLMIR,boink2019partially}, even if their performance is shown to be quantitatively and qualitatively superior \cite{hsu2021comparing}. This is primarily due to the extensive computational cost of evaluating the forward model, which limits reconstruction speeds and often makes training prohibitive. Thus, the majority of learned approaches in PAT consider either one or two-step approaches, that is a direct mapping from data to reconstruction is learned or a post-processing network after an initial reconstruction is obtained, respectively. We refer to \cite{hauptmann2020deep,grohl2021deep,yang2021review} for a comprehensive overview of learning based reconstruction methods for PAT.

As previously discussed, a possibility to overcome the limitations of the expensive forward model in the training procedure for PAT has been proposed in \cite{Hauptmann2018}, by training the update operators in a greedy approach only requiring iterate wise optimality. The use of faster approximate models for planar (or linear) sensors \cite{Koestli2001,Cox2005} further speeds up training and deployment times \cite{Hauptmann2018}. An alternative route considers to learn the forward model entirely, for instance by Fourier Neural Operators \cite{hsu2023fast,guan2023fourier} to embed these into an efficient iterative reconstruction algorithm.

\subsection{A fast approximate forward and inverse model}\label{sec:fastModels}
Recent studies have investigated the effect of using approximate or imperfect operators in the reconstruction \cite{bungert2020variational,arjas2023sequential,nicholson2023global} and the possibility to improve results with a learned correction \cite{lunz2021learned,smyl2021learning}. Most importantly, it is shown that under a suitable learned correction, one can accurately solve the classic variational problem. We will first discuss the approximate model here and then introduce the learned reconstruction later in sec. \ref{sec:MC-PD}.

A fast forward and backward model is needed to enable feasible training of a learned iterative reconstruction. Here we will use an analytical approximation based on the Fast Fourier Transform (FFT), that is applicable for line and planar sensors \cite{Koestli2001,Cox2005}. We note, that an FFT based algorithm also exists for circular domains \cite{kunyansky2012fast} and thus our method extends naturally. We shortly review our approximate model as previously used in \cite{hauptmann2018MLMIR}.

The approximate model exploits the fact that the measurement points lie on a line in 2D at $\mathbf{x}_2=0$ (or plane in 3D), and assuming constant sound-speed $c$. In two dimensions the pressure on the sensor can be related to $x$ by \cite{Cox2005,Koestli2001}:
\begin{align}
p(\mathbf{x}_1,t) = \frac{1}{c^2} 
\mathcal{F}_{k_1}\left\{\mathcal{C}_{\omega}\left\{
B(k_1,\omega) \tilde{x}(k_1,\omega)
\right\}\right\},
\label{eqn:FastFwd}
\end{align}
where 
$\tilde{x}(k_1,\omega)$ is obtained from $\hat{x}(\mathbf{k})$ via the dispersion relation $(\omega/c)^2 = k_1^2+k_2^2$ and 
$\hat{x}(\mathbf{k}) = \mathcal{F}_{\mathbf{x}}\{x(\mathbf{x})\}$ is the 2D Fourier transform of $x(\mathbf{x})$. $\mathcal{C}_{\omega}$ is a cosine transform from $\omega$ to $t$, and $\mathcal{F}_{k_1}$ is the 1D inverse Fourier Transform on the detector line. Finally, the weighting factor
\vspace{-0.5em}
\begin{align}
B(k_1,\omega) = \omega/\left(\textrm{sgn}(\omega)\sqrt{(\omega/c)^2 - k_1^2 }\right),
\end{align}
contains an integrable singularity which means that if \eqref{eqn:FastFwd} is evaluated by discretization on a rectangular grid to enable the application of the FFT for efficient calculations, then aliasing in the measured data $p(\mathbf{x}_1,t)$ results. Consequently, evaluating \eqref{eqn:FastFwd} using FFT leads to a fast but approximate forward model. One could control the strength of aliasing artifacts by thresholding certain incident angles as in  \cite{hauptmann2018MLMIR}, but this leads to a loss of signal strength. Thus, here we will not use a thresholding and deal with the aliasing by the learned model correction.

The relation in \eqref{eqn:FastFwd} then defines the approximate forward model $\widetilde{A}:X\to Y$ for this study.
Similarly, we can invert \cref{eqn:FastFwd} to obtain a mapping from measured time-series to initial pressure as inverse mapping or reconstruction operator $A^\dagger:Y\to X$. This leads to two mappings between image $X$ and data $Y$ space: 
\begin{equation}\label{eqn:approxOps}
\widetilde{A} x = \tilde{y}, \quad A^\dagger y = \tilde{x}.
\end{equation}
We note, that the fast inverse mapping does not introduce any aliasing, but information is naturally limited by the information present from the limited-view geometry, see Figure \ref{fig:dataGeneration} for an illustration. In the following we will use the above fast mappings in the learned reconstruction. 

Additionally, we have implemented the mappings in eq. \eqref{eqn:approxOps} with PyTorch to enable flexible deployment in the training of the unrolled algorithms. This way, we have full GPU and automatic differentiation support of the models, enabling efficient training and evaluation. As part of this study, we will make our implementation of the models available to researchers.

\section{Model corrected primal-dual and deep equilibrium models}\label{sec:MC-PD}
We can now combine the above pieces  to formulate a learned iterative scheme starting from the classic PDHG formulation \eqref{eqn:PDHG-dual} and \eqref{eqn:PDHG-primal}. We will first include the model correction term for the forward operator and include a learnable update operator in the primal space. Additionally, our formulation allows to formulate the proposed algorithm as a deep equilibrium model. 
%This gives raise to the here proposed model corrected primal-dual models for limited-view photoacoustic tomography.

\subsection{Model corrected learned primal dual}
The plug-and-play framework suggests to substitute the proximal operator in \cref{eqn:proxgrad} with a learned network. This is known to work well, if restricted to the primal space. The same approach is usually not applied to learning the proximal operator for the dual space, as we can not simply learn a denoising network in the dual space. 
Additionally, the proximal operator in dual space has a clear connection to the data-fidelity term and the noise model. Thus, for our method we will keep the explicit form of the proximal in dual space as given by \cref{eqn:PDHG-dual} and only replace the proximal operator for the primal update in image space with a network $G_\theta$.

We replace the accurate but expensive model $A$ with a fast approximate version $\widetilde{A}$, which would lead to errors in the data and hence the dual update. As solution, we propose to learn a correction of the forward operator by a network $F_\phi$. 
%, such that $F_\phi(\widetilde{A}x)\approx Ax$ in the statistical sense, i.e., for a class of relevant $x$ (defined by the training data). 
Instead of using the (expensive) adjoint in the primal space, we will make use of the fast inverse mapping $A^\dagger$. Together, the update steps with the model correction and learned proximal operator are then given by
\begin{align}
    q_{k+1} &=\frac{q_k + \sigma (F_\phi(\widetilde{A}x_k) - y)}{1+\sigma} \label{eqn:MC-PD-dual} \\
    x_{k+1} &=G_\theta\left(x_k - \tau A^\dagger q_{k+1}\right). \label{eqn:MC-PD-primal}
\end{align}
Note, that we use weight sharing of the network parameters $\theta,\phi$ for all iterates, and we set a maximum number of iterations $K>0$.  We refer to this unconstrained version as model-corrected learned primal-dual (MC-PD) as outlined in Algorithm \ref{alg:MCPD}.

\begin{algorithm}
	\caption{Model-Corrected Primal-Dual (MC-PD)}
	\label{alg:MCPD}
	\begin{algorithmic}[1]
	    \State Given $y,\widetilde{A},A^\dagger$
        \State Set $\sigma, \tau, K$
		\Let{$x_0$}{$A^\dagger y$}
		\Let{$q_0$}{$0$}
        \Let{$k$}{$0$}
        \For{$0<k\leq K-1$} 
        \Let{$q_{k+1}$}{$\frac{q_k + \sigma (F_\phi(\widetilde{A}x_k) - y)}{1+\sigma}$}
        \Let{$x_{k+1}$}{$G_\theta\left(x_k - \tau A^\dagger q_{k+1}\right)$} 
        \EndFor
	\end{algorithmic}
\end{algorithm}

\subsection{A Primal-dual deep equilibrium model}
In recent years convergent learned image reconstruction methods have gained considerable interest. Instead of simply training the learned reconstruction in a supervised method, we can constrain weights to obtain theoretical guarantees on the behavior of the unrolled iterates. One such option is to require that the learned iterates converge to a fixed point, this is the underlying paradigm of deep equilibrium models. In its simplest form, we can ensure that the learned proximal gradient in \cref{eqn:learnedProx} satisfies a fixed-point equation, that is
\begin{equation}\label{eqn:DEQ}
    x^*= G(x^*;y) =  G_\theta(x^* - \lambda A^*(Ax^*-y)).
\end{equation}
First we note, that weights are shared between iterates and no maximum number of iterations is assumed, but rather a convergence in the limit is considered by ensuring, e.g, that the updates are a contraction. In deep equilibrium models the forward solver applies a fixed-point iteration, either a vanilla version or more advanced choices such as Anderson acceleration, see \cite{bai2019deep} for details and an extension to inverse problems in \cite{gilton2021deep}. Training is then performed by differentiating the fixed-point formulation, eliminating the need to perform backpropagation through all iterates and hence reducing memory consumption.

In our case, we are given the two variables $x,q$ for which we would like to formulate the deep equilibrium model. That means, we are now looking for a pair of fixed points $z^*=(x^*,q^*)\in X\times Y$, such that the fixed-point equation is satisfied with
\begin{equation}\label{eqn:PD-DEQ}
    (x^*,q^*)=\mathrm{PD}_{(\phi,\theta)}(x^*,q^*;y),
\end{equation}
where the corresponding update equations for the mapping $\mathrm{PD}$ are given by \cref{eqn:MC-PD-dual} and \cref{eqn:MC-PD-primal}.  In the following we will first discuss the forward pass, followed by a discussion on the computation of backpropagation. 

\subsubsection{Forward pass for the deep equlibrium formulation}
Here, we will be applying the Anderson accelerated fixed-point method to both, primal and dual, variables. More precisely, given the primal-dual update as above for iterate $k$:\[
    (x_{k+1},q_{k+1})=\mathrm{PD}_{(\phi,\theta)}(x_k,q_k;y).
\]
%defined by the equations \eqref{eqn:MC-PD-dual} and \eqref{eqn:MC-PD-primal}. 
We denote with a superscript which update we compute, i.e., $\mathrm{PD}^q$ and $\mathrm{PD}^x$, that is we get the formulation
\begin{align}
    q_{k+1}&=\sum_{i=0}^{m-1} \beta_i\mathrm{PD}^q_{(\phi,\theta)}(x_{k-i},q_{k-i};y), \label{eqn:AndersonDual}\\
    x_{k+1}&=\sum_{i=0}^{m-1} \alpha_i\mathrm{PD}^x_{(\phi,\theta)}(x_{k-i},q_{k-i};y), \label{eqn:AndersonPrimal}
\end{align}
with memory $m\geq 1$, $\sum_i\alpha_i=1$ and $\sum_i\beta_i=1$. The parameter vectors $\mathbf{\alpha}$ and $\mathbf{\beta}$ are computed by solving
\[
\min_\alpha \|H\alpha\|_2^2, \text{ subject to } \sum_i\alpha_i=1, 
\]
where the matrix $H$ consists of the vectorized residuals
\[ 
\begin{split}
H=\left[\mathrm{PD}^x_{(\phi,\theta)}\right.&(x_k,q_k;y) - x_k, \cdots,  \\  &\left.\mathrm{PD}^x_{(\phi,\theta)}(x_{k-m+1},q_{k-m+1};y) - x_{k-m+1}\right],
\end{split}
\]
and similarly we perform the same for $\beta$ and the dual variable $q$. The forward pass of the primal-dual deep equilibrium model (PD-DEQ) is then given similarly as in Algorithm \ref{alg:MCPD}, except that new iterates are computed by the Anderson accelerated version in eqs. \eqref{eqn:AndersonDual} and \eqref{eqn:AndersonPrimal}. Additionally, one could not only consider a maximum number of iterations but also check if the fixed-point iteration has converged, for instance, by setting a limit on the relative residual of the primal variable 
\[
r=\frac{\|PD^x_{(\phi,\theta)}(x_{k+1},q_{k+1};y)-x_k\|_2}{\|PD^x_{(\phi,\theta)}(x_{k+1},q_{k+1};y)\|_2}.
\]
% \begin{algorithm}
% 	\caption{Primal-Dual Deep Equilibrium Model (PD-DEQ)}
% 	\label{alg:PD-DEQ}
% 	\begin{algorithmic}[1]
% 	    \State Set $\sigma, \tau, K, \delta$ (more?)
% 		\Let{$x_0$}{$G_\vartheta(A^\dagger y)$}
% 		\Let{$q_0$}{$0$}
%         \Let{$k$}{$0$}
%         \While{$k\leq K$ \& $r>\delta$} 
%         \Let{$(x_{k+1},q_{k+1})$}{DEQ update $\mathrm{PD}_{(\phi,\theta)}(x_k, q_k;y)$}
% %        \Let{$x_{k+1}$}{DEQ update $\mathrm{PD}_{(\phi,\theta)}(x_k, q_k;y)$}
%         \Let{$r$}{$\frac{\|PD^x_{(\phi,\theta)}(x_{k+1},q_{k+1};y)-x_k\|_2}{\|PD^x_{(\phi,\theta)}(x_{k+1},q_{k+1};y)\|_2^2}$}
%         \Let{$k$}{$k+1$}        
%         \EndWhile
% 	\end{algorithmic}
% \end{algorithm}

\subsubsection{Backward pass}

For training, even though we perform the update on both variables, we only optimize the network parameters with respect to a loss on the primal variable $x$, where training data is easily available. That means, we formulate the fixed-point equation in \eqref{eqn:PD-DEQ} only with respect to $x$. We write for a fixed point $x^*$ under dual variable $q$ and measurement $y$, with a slight abuse of notation, 
\begin{equation}\label{eqn:fixedPointEq}
    x^*=\mathrm{PD}_{(\phi,\theta)}(x^*;q,y).
\end{equation}
To simplify notation we denote the network variables as one $\zeta=\{\phi,\theta\}$. The networks are then trained supervised to minimize the mean-squared error (MSE) with respect to the groundtruth $x$ by
\[
\ell(\zeta) = \| \mathrm{PD}_{(\zeta)}(x^*;q,y) - x\|_2^2 = \| x^* - x\|_2^2.
\]
The backward pass is computed by implicit differentiation of the fixed-point equation \eqref{eqn:fixedPointEq}. We denote the Jacobian at the fixed point with respect to the parameters $\zeta$ by $J_{\mathrm{PD}^{x^*}}(\zeta)$ and the Jacobian with respect to $x^*$ by $J_{\mathrm{PD}_{(\zeta)}} (x^*)$, then we obtain for the gradients the implicit equation
\begin{equation}\label{eqn:implicitDiff}
    \nabla\ell(\zeta) = \bigl(J_{\mathrm{PD}^{x^*}}(\zeta)\bigr)^T\bigl( \mathrm{Id} - J_{\mathrm{PD}_{(\zeta)}} (x^*) \bigr)^{-T}\bigl(x^*-x\bigr). 
\end{equation}
Note, that we do not need to compute the backwards pass through all iterates and thus memory consumption is independent of the number of unrolled iterates. Nevertheless, eq. \eqref{eqn:implicitDiff} requires a solver for the implicit equation, which can add some computational overhead. We will consider here the full formulation, but further reductions in the computation can be obtained by considering approximations of $J_{\mathrm{PD}_{(\zeta)}} (x^*)$, most notably a Jacobian free variant \cite{fung2022jfb}.

While our primary interest lies in the convergence of the primal $x$ it is now reasonable to ask what do we need to assume on $q$. Let us discuss this in the following.

\subsection{Convergence analysis}\label{sec:convAna}
Here we will shortly examine some theoretical properties of the proposed methods. We investigate how the model correction can be interpreted in the MC-PD formulation. Followed by a discussion on the contraction and fixed-point properties of the deep equilibrium model.
\subsubsection{Functional convergence under forward correction}
We start by considering the analytic PDHG formulation in eqs. \eqref{eqn:PDHG-dual}-\eqref{eqn:PDHG-relaxation} using the approximate operator $\widetilde{A}$, its corresponding adjoint $\widetilde{A}^*$, and the proximal operator corresponding to some regularization $R$. Given the operator norm $\widetilde{L}=\|\widetilde{A}\|$, we can choose the hyperparameters as $\sigma=\tau<1/\widetilde{L}$ and $\alpha\in[0,1]$. Then, applying PDHG will minimize the cost functional 
\[
\|\widetilde{A}x-y\|_2^2 + \lambda R(x).
\]
Minimizing the above functional with the approximate forward operator instead of the correction does result in suboptimal results as demonstrated in \cite{lunz2021learned}. 

Consequently, let us now consider including the forward operator correction in the dual update, i.e., we replace $A$ with $F_\phi(\widetilde{A})$ in eqn. \eqref{eqn:PDHG-dual}, while using the correct adjoint in the primal update \eqref{eqn:PDHG-primal}. We denote the operator norm of the correction $F_\phi$ by $\varepsilon_\phi = \|F_\phi\|$. Then, under a perfect training assumption on the adjoint we can recover convergence of PDHG as summarized in the following result. 
\begin{proposition}\label{theo:convCorr}
Given corrected forward operator $F_\phi(\widetilde{A})$ with norm $\varepsilon_\phi \widetilde{L}$, accurate adjoint $A^*$ with norm $L$. Assume that for $k\geq 0$ we have
\begin{equation}\label{eqn:adjointFit}
    \bigl(F_\phi(\widetilde{A}x_k),y_k\bigr)_Y=\bigl(x_k,A^*y_k\bigr)_X.
\end{equation}
Then for the choice $\tau< 1/L$ and $\sigma < 1/(\varepsilon_\phi \widetilde{L})$ we recover the iterations of PDHG in eqs. \eqref{eqn:PDHG-dual}-\eqref{eqn:PDHG-relaxation} for the accurate functional \eqref{eqn:variational}.
\end{proposition}
\begin{proof}
The condition \eqref{eqn:adjointFit} requires that the forward correction does satisfy the adjoint equation. By uniqueness of the adjoint this is equivalent to a perfect training condition
\[
\bigl(F_\phi(\widetilde{A}x_k)-Ax_k,y_k\bigr)_Y = 0 \, \hspace{0.25cm } \forall k\geq 0.
\]
Thus, all iterates in \eqref{eqn:PDHG-dual}-\eqref{eqn:PDHG-relaxation} are identical to the accurate operators. By the choice of hyper parameters $\sigma$ and $\tau$ we have that $\tau\sigma L \widetilde{L}\varepsilon_\phi < 1$ and thus the the convergence condition of the classic PDHG is satisfied.
\end{proof}

The condition \eqref{eqn:adjointFit} could be used for training without the need to access the accurate forward operator. That is one could add an additional adjoint loss to the training 
\begin{equation}\label{eqn:adjointLoss}
L(x;\phi)={(F_\phi(\tilde{A}x),h)_Y - (x,A^* h)_X} \text{ for } h\in Y.
\end{equation}
Clearly, a perfect training can not be achieved in practice. In this case, it can be treated as a perturbation in the adjoint, or, in fact, in the forward. We refer to \cite{lorenz2023chambolle,chouzenoux2023convergence} for a discussion on convergence under mismatched adjoints in primal-dual algorithms.

Note, that in the above adjoint loss \eqref{eqn:adjointLoss} the accurate adjoint is still needed though. Thus, we now consider the case, where both forward and adjoint are replaced by their respective fast operators. Then, the corresponding adjoint loss becomes
\begin{equation}\label{eqn:adjointLoss_inverse}
L(x;\phi)={(F_\phi(\tilde{A}x),h)_Y - (x,A^\dagger h)_X} \text{ for } h\in Y,
\end{equation}
where adjointness of the forward correction to the fast inverse is enforced. Clearly, this does not minimize the accurate functional, but rather the corresponding functional 
%Clearly, the above case does not reflect our algorithm as we still need access to the computationally expensive adjoint. We can adjust the result by replacing $A^*$ with the fast inverse $A^\dagger$ and norm $L^\dagger$. 
\[
\|F_{\phi}(\widetilde{A})x-y\|_2^2 + \lambda R(x),
\]
where closeness to the true model is governed by the closeness of the fast inverse $A^\dagger$ to the adjoint $A^*$. If convergence with respect to the accurate functional is desired, an adjoint correction is required \cite{lunz2021learned}.

\subsubsection{fixed-point convergence for the deep equilibrium model}
Let us now replace the proximal operator with a learned network $G_\theta$ and corresponding norm $\varepsilon_\theta$. First we note, that if we do not enforce any further conditions on $G_\theta$, we can not expect functional convergence anymore, but we can obtain a fixed-point convergence \cite{mukherjee2023learned}. 
We remind first, that training of the PD-DEQ model is performed only with respect to the primal variable $x$. Thus, we would like to verify two conditions 
\begin{itemize}
    \item[(i)] Under which parameter choices and conditions on the networks $F_\phi$ and $G_\theta$ do we obtain a contraction on the primal variable?
    \item[(ii)] Can we also ensure fixed-point convergence?
\end{itemize}
As we will see, both conditions are closely connected as condition (ii) will require a contraction, but we will see that a dependence on $q$ will not provide fixed-point convergence without further assumptions. To begin with, let us first combine the update equations \eqref{eqn:MC-PD-dual} and \eqref{eqn:MC-PD-primal} by 
\[
x_{k+1}=G_\theta\Bigl(x_k-\tau A^\dagger\bigl( q_k+\sigma\bigl(F_\phi(\widetilde{A}x_k) - y\bigr)\bigr) / (1 + \sigma)\Bigr).
\]
We can see here the dependence on $q_k$ in the updates. Let us isolate the update for $x$ by introducing the operator $T:X\to X$ such that
\begin{equation}\label{eqn:primal_update_T}
T_{\mathrm{Id}}(x) := (\mathrm{Id} - T)(x) =  x - \tau A^\dagger \sigma F_\phi(\widetilde{A}x)/(1+\sigma).
\end{equation}
Thus, $T$ is a scaled version of the normal operator $A^*A$. We note, the above update rule on $x$ \eqref{eqn:primal_update_T} convergences to a fixed-point if $T$ is firmly nonexpansive, which implies that $\mathrm{Id} - T$ is firmly nonexpansive \cite{bauschke2012firmly}.
To show firmly nonexpansiveness we need an adjoint loss condition. Rather than assuming a perfect fit, we consider an approximate adjoint loss condition, such that
\begin{equation}\label{eqn:approximateAdjointLoss}
    (F_\phi(\tilde{A}x),h)_Y \leq (1+\varepsilon)(x,A^\dagger h)_X \text{ for } h\in Y.
\end{equation}
That means we only fulfill the adjoint condition with an $\varepsilon>0$ error depending on the value of the right inner product. This is a more realistic case, as we can not expect a perfect adjoint fit. We note, that the assumption \eqref{eqn:approximateAdjointLoss} requires $(x,A^\dagger h)_X$ to be positive. However, since all samples $x$ and relevant $h$ are positive we can assume this to hold. 
Now we can show the firmly nonexpansiveness of $T$. 
\begin{theorem}\label{theo:nonexpansive}
The operator $T:X\to X$ defined in \eqref{eqn:primal_update_T} is firmly nonexpansive, under condition \eqref{eqn:approximateAdjointLoss} with $\varepsilon>0$, $L^\dagger=\|A^\dagger\|$, and $\sigma,\tau>0$ chosen such that
\begin{equation}\label{eqn:nonexpansiveCoefCond}
    (1+\varepsilon)\frac{\tau\sigma}{1+\sigma}(L^\dagger)^2 \leq 1. 
\end{equation}
\end{theorem}
\begin{proof}
%First, we note that the normal operator $\lambda A^*A$ is firmly nonexpansive, if $\lambda\|A^*A\|=\lambda\|A\|^2\leq 1$. 
We need to show that
\begin{equation}\label{eqn:nonexpansiveCond}
\|T(x)-T(v)\|_2^2 \leq (T(x)-T(v),x-v)_X,
\end{equation}
where $T:X\to X$ is given as in \eqref{eqn:primal_update_T}. 
We start by bounding the left hand side of the above inequality, using linearity of $A^\dagger$ and $\|A^\dagger\|=L^\dagger$, we get
\begin{align*}
    \Bigl\|&\frac{\tau\sigma}{1+\sigma} A^\dagger F_\phi(\widetilde{A}x) - \frac{\tau\sigma}{1+\sigma} A^\dagger F_\phi(\widetilde{A}v)\Bigr\|_2^2  \\  
    & \leq  \Bigl|\frac{\tau\sigma}{1+\sigma}\Bigr|^2
     \Bigl|L^\dagger\Bigr|^2  \Bigl\|F_\phi(\widetilde{A}x) - A^\dagger F_\phi(\widetilde{A}v)\Bigr\|_2^2  \\
     &= \Bigl|\frac{\tau\sigma L^\dagger}{1+\sigma}\Bigr|^2
     \Bigl(F_\phi(\widetilde{A}x) - F_\phi(\widetilde{A}v),F_\phi(\widetilde{A}x) - F_\phi(\widetilde{A}v) \Bigr)_X. 
\end{align*}
We can now separate the inner product, using semi-linearity and symmetry. Then we estimate each term using the approximate adjoint condition in \eqref{eqn:approximateAdjointLoss} and recombine, to obtain 
\begin{align*}
     \Bigl(&F_\phi(\widetilde{A}x) - F_\phi(\widetilde{A}v),F_\phi(\widetilde{A}x) - F_\phi(\widetilde{A}v) \Bigr)_X \\
      &\leq (1+\varepsilon)\Bigl(A^\dagger F_\phi(\widetilde{A}x) - A^\dagger F_\phi(\widetilde{A}v),x-v \Bigr)_X. 
\end{align*}
Now under the condition \eqref{eqn:nonexpansiveCoefCond} we get the required estimate.
\end{proof}

We continue by examining the iterations of the learned primal-dual updates and obtain a contraction for fixed $k$ under boundedness of the dual updates.

\begin{theorem}\label{theo:contraction}
Assume $\|q_k-q_{k-1}\|_2<C_q$ with $k>0$ fixed and $\|G_\theta\|=\varepsilon_\theta<1$. Under the assumptions in Theorem \ref{theo:nonexpansive}, there exists a choice for $\tau>0$, such that the iteration for $x_{k+1}$ is contracting, i.e., 
\[
\|x_{k+1}-x_k\|_2 < \|x_{k}-x_{k-1}\|_2.
\]
\end{theorem}
\begin{proof}
We start by reformulating the update for $x_{k+1}$ using $T_\mathrm{Id}$ as
\[
x_{k+1}=G_\theta\Bigl( T_\mathrm{Id}(x_k) - \frac{\tau}{(1+\sigma)} A^\dagger( q_k - \sigma y) \Bigr ). 
\]
With the above update rule, using the norm of the network $\|G_\theta\|=\varepsilon_\theta$ and $\|A^\dagger\|=L^\dagger$ we can obtain by triangle inequality a first bound of iterates as
\begin{equation}\label{eqn:theo_iterBound}
\begin{split}
\|x&_{k+1}-x_k\|_2\leq \\ &\varepsilon_\theta \Bigl( \bigl\|T_\mathrm{Id}(x_k)-T_\mathrm{Id}(x_{k-1})\bigr\|_2 + \frac{\tau L^\dagger}{1+\sigma}\bigl\|q_k-q_{k-1}\bigr\|_2\Bigr).
\end{split}
\end{equation}
Set $0<\delta = 1-\varepsilon_\theta<1$, with $\|q_k-q_{k-1}\|_2<C_q$ and since $ \|x_k-x_{k-1}\|_2$ is finite, there exists $\tau>0$ small enough such that 
\begin{equation}\label{eqn:theo_tauEstimate}
\frac{\tau L^\dagger}{1+\sigma}C_q < \delta/2 \|x_k-x_{k-1}\|_2.
\end{equation}
We then obtain the final estimate for \cref{eqn:theo_iterBound}, using firmly nonexpansiveness of $T_\mathrm{Id}$ in the first term and \eqref{eqn:theo_tauEstimate} for the second term, by
\[
\begin{split}
\|x_{k+1}-x_k\|_2 &\leq \varepsilon_\theta(1 + \delta/2)\|x_k-x_{k-1}\|_2 \\
&< \|x_k-x_{k-1}\|_2,
\end{split}
\]
since by assumption
\[
\varepsilon_\theta(1+\delta/2)<\varepsilon_\theta+\delta/2=(1+\varepsilon_\theta)/2<1.
\]
\end{proof}
The above theorem only provides an iterate-wise convergence under a uniform bound on the dual $q$. That means, if $q_k$ is not converging $\tau$ may become arbitrarily small. We note, that in practice we only consider finite iterates and hence a uniform bound exists. For the limit case we could assume that $q_k$ converges at least as fast as $x_k$, then we can obtain a uniform choice for $\tau>0$. For that, let us assume that both variables have a $1/k$ convergence.
\begin{corollary}
    Under the assumptions of Theorem \ref{theo:contraction}, assume additionally that both $\|q_k-q_{k-1}\|=\mathcal{O}(1/k)$ and $\|x_k-x_{k-1}\|=\mathcal{O}(1/k)$. Then there exists a uniform $\tau>0$ for all $k>0$ such that the $x_{k}$ are contracting.
\end{corollary}
\begin{proof}
By assumption both sides of the estimate \cref{eqn:theo_tauEstimate} can be divided by $k$ (with possible additional constants on each side). Then there exists a small enough $\tau>0$ that holds for all $k$. 
\end{proof}
First of all, the above analysis reveals that it is advisable to choose $\tau$ and $\sigma$ conservatively small to ensure that the iterations contract. As we will see, this is in fact important to have stability in the DEQ formulation.
Secondly, we would like small norm of the residuals in the dual $q$ and ideally fast convergence. Third, it is interesting to note that in fact only the dual network $G_\theta$ is required to be contractive, while the model correction network $F_\phi$ is connected to the inverse $A^\dagger$ by the approximate adjoint error. 
In the experiments, we will see that $q$ does indeed converge fast and the residuals have significantly smaller norm than for $x$.
Thus, the experiments confirm that we are able to recover a fixed-point iteration on both primal and dual variables, even if the theory assumes dependence on the contraction of $q$.

\section{Photoacoustic data simulation and computational aspects}\label{sec:CompMethods}
In this study we will evaluate the performance of the proposed algorithms with a simulated two-dimensional scenario.
For this, we will first discuss the training data generation. We will then continue to discuss some details on the implementation of forward and inverse models, followed by the employed training procedures.

\subsection{Photoacoustic simulation in 2D and training data}
\label{sec:2Ddata}
For simulating training data we use k-wave, which is based on a pseudo-spectral method, for accurate simulation of the acoustic wave equation \eqref{eqn:PATfwd} in two dimensions. We consider a rectangular computational domain of size $80 \times 128$ with isotropic spatial pixel size of $dx = 106 \mu$m. The photoacoustic sensor is located at the top edge of the domain, see Figure \ref{fig:dataGeneration} for an illustration. The sound speed is set to be constant at 1500 m/s and temporal sampling rate at $dt = 50$ ns, the measurement data is of dimension $160\times 128$, with the spacing $dt \times dx$. We have added 1\% Gaussian random noise to the simulated measurements, which corresponds to a SNR of roughly 18dB.

For training data generation we used the \emph{DRIVE: Digital Retinal Images for Vessel Extraction}\footnote{https://drive.grand-challenge.org/}, consisting of retinal photographs and segmentation masks for the training. We have used 20 images and corresponding masks to create phantoms of vessel structures. First, by converting the images to grayscale and normalization, then by multiplying the mask with the images. Each resulting image has dimension $584\times 565$, we have then extracted smaller patches of size $80\times 128$ from the images and only saved those with enough features by checking the sum over all pixels, i.e., if greater than 150. We have repeated the same procedure for each image transposed for data augmentation, this resulted in a total of 893 images, we have further augmented this data set by flipping the image vertically, since this will provide different limited view artifacts. In total we obtained 1786 images with a split into $N=1600$ for training and 93 for validation and testing each.

\subsection{Implementing the fast models}
For this study, we have implemented both models, fast approximate forward $\widetilde{A}$ and the fast inverse $A^\dagger$, fully in PyTorch, to allow for efficient evaluation. This also enables to use automatic differentiation within the model-based learned reconstructions. The implementation follows largely the description in sec. \ref{sec:fastModels} using the inbuilt FFT with PyTorch. For the mapping between image $k$-space, $\mathbf{k}=(k_1,k_2)$, and measurement $k$-space, $(k_1,\omega)$, we have implemented our own grid interpolator within PyTorch, as no inbuilt functions were available. 

For efficient computation during training, all necessary static variables and $k$-grids are pre-computed and passed to the fast model. This follows a similar paradigm as in the $k$-wave toolbox \cite{treeby2010}. 
As part of this study we publish the fast models and network codes. \footnote{Link will be added after revision}

\subsection{Implementation and training of the model-corrected primal dual models}\label{sec:DEQ-train}
The networks $F_\phi$ and $G_\theta$ for the proposed models are both chosen as U-Nets, with 3 scales, i.e., 2 downsampling layers, 64 channels in the first scale, followed by 128 and 256 in the next scales. 
We set a maximum of 10 iterations for both algorithms, the Anderson acceleartion of PD-DEQ uses a memory of $m=5$. For the hyperparameters $\tau$ and $\sigma$ we have computed an approximate operator norm of $\widetilde{L}\approx 0.7$ and set $\tau=\sigma=1/(10 \widetilde{L})$. The overestimation by a factor 10 is due to the presented theory and provided improved stability in practice.

As mentioned earlier, we only train the networks with respect to the primal variable. We are given ground-truth images $x^i$ and corresponding measurements $y^i$ and the corresponding reconstruction operator $\mathcal{R}_\zeta:Y\to X$ (either MC-PD or PD-DEQ) with parameters $\zeta=\{\phi,\theta\}$. We then minimize the $\ell^2$-loss (MSE) to find the optimal set of parameters $\zeta^*$ as
\begin{equation}\label{eqn:loss}
    \zeta^* = \argmin_\zeta \sum_{i=1}^N \|\mathcal{R}_\theta(y^i) - x^i \|_2^2.
\end{equation}
We only use 1 sample per training iteration due to memory constraints and at each iteration the training sample is drawn from the uniform distribution over all training pairs. We use the Adam optimizer in PyTorch with a cosine decay on the learning rate initialized by $2\cdot 10^{-4}$ and a total of 25000 training iterations. Training of the MC-PD models took 100 minutes and 180 minutes for the PD-DEQ models on a Nvidia Quadro RTX 6000 with 24 GB memory.

For the training of the PD-DEQ model, we use automatic differentiation at the fixed point formulation to compute the gradients given by eq. \eqref{eqn:implicitDiff}. We note that solving eq. \eqref{eqn:implicitDiff} stably requires the network to be contractive. Thus, we have tested to use spectral normalization to explicitly constrain the Lipschitz constant of $G_\theta$, but better results have been obtained without, we refer to Section \ref{sec:spectral} for a discussion. This suggests that the contraction is implicitly enforced by the fixed-point formulation.
Nevertheless, we remind that a Lipschitz constraint is only needed for the update network $G_\theta$ in primal space, while the dual network $F_\phi$ can be unconstrained. We also point out that supervised training with respect to the dual variable is not straight-forward as the ideal reference should only be residual noise.

\subsubsection{A combined hybrid approach}

We will see that the PD-DEQ model has some difficulties to achieve high values of PSNR in the presented limited-view setting, despite providing good image quality. Thus, we will additionally consider a hybrid combination of both approaches, MC-PD and PD-DEQ. That means we first perform a fixed amount of MC-PD iterations, followed by PD-DEQ. The order is chosen such that first MC-PD will help to alleviate the impact of the limited-view setting, then the final PD-DEQ iterations provide the fixed-point formulation for to the final reconstructions. Here, we need to find a trade-off between the MC-PD performance and the memory reduction provided by PD-DEQ, as we will discuss later. We call this approach in the following PD-Hybrid

For the training objective, we train the network in an alternating way, that is in each training iteration we first evaluate MC-PD, denoted by $\mathcal{R}_{\zeta_1}^{MC}$ and minimize \eqref{eqn:loss} for sample $i$. Then given the output $(x_k^i,q_k^i)=\mathcal{R}_{\zeta_1}^{MC}(y^i)$ we evaluate the PD-DEQ, denoted by $\mathcal{R}_{\zeta_2}^{DEQ}$, and minimize the loss function only for the parameters of the PD-DEQ model, i.e.,
\[
\|\mathcal{R}_{\zeta_2}^{DEQ}(x_k^i,q_k^i,y^i) - x^i \|_2^2.
\]
We will use 5 iterations of MC-PD followed by 5 iterations of PD-DEQ. Training of the PD-Hybrid models takes about 160 minutes.

{\newcommand{\showpic}[3]{%
\begin{tikzpicture}%[spy using outlines={circle, magnification=4, size=2cm, connect spies}]%

\draw (0,0) node [anchor=north] {\includegraphics[width=6cm]{recons/8_#3}};%
\draw (0,0) node [anchor=south, yshift=-10pt] {\phantom{f}#1\phantom{g}};%
\draw (0,0) node [anchor=south, yshift=-20pt] {\phantom{f}#2\phantom{g}};
%\spy on (0,-1.15cm) in node at (-.2cm, -3.5cm);
\end{tikzpicture}\hspace*{-2mm}%
}

\begin{figure*}[t!]
\centering
\showpic{Phantom}{}{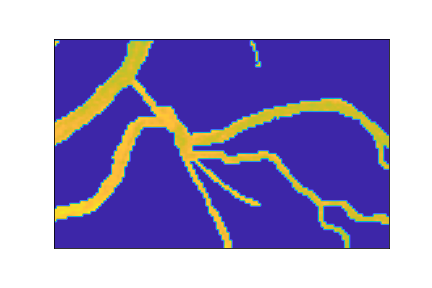}%
\hfill
\showpic{U-Net}{PSNR: 20.07 / SSIM: 0.946}{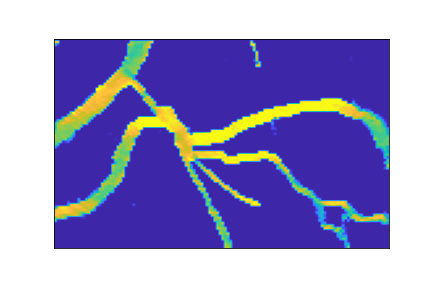}%
\hfill
\showpic{LPD}{PSNR: 26.23 / SSIM:  0.987}{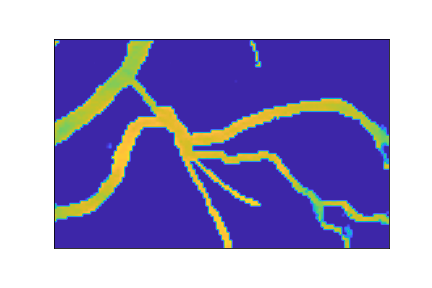}%
\\ \vspace{-0.5cm}
\showpic{MC-PD}{PSNR: 27.23 / SSIM: 0.991}{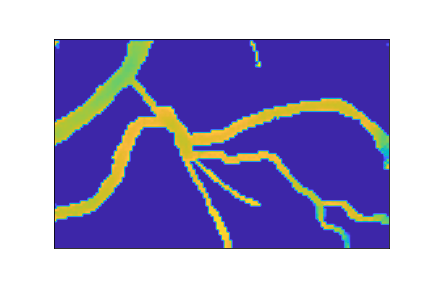}%
\hfill
\showpic{PD-DEQ}{PSNR: 21.64 / SSIM: 0.962}{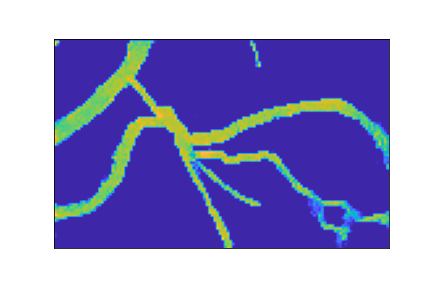}%
\hfill
\showpic{Hybrid}{PSNR: 23.87 / SSIM: 0.979}{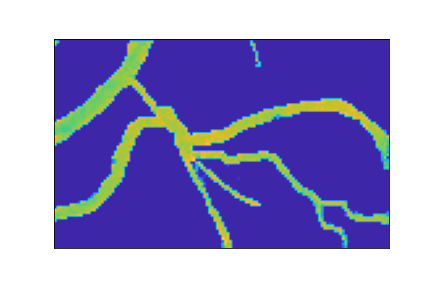}%
\\ \vspace{-0.5cm}
\includegraphics[width = 0.8\linewidth]{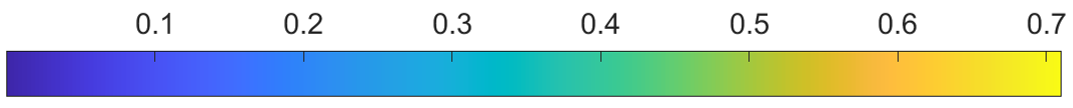}
\caption{\label{fig:recons_test} Obtained reconstructions for the considered methods on a sample from the test set with quantitative measures. All images are on the same color scale.  Measurements are taken on the top edge, with 1\% Gaussian noise. The resolution of reconstructions is $80\times 128$.}
\end{figure*}}

\subsection{Comparison methods}

We will further compare the performance of both model-corrected primal-dual models with two well-established methods. First we use a post-processing approach, where an initial reconstruction is obtained with the inverse $x_0=A^\dagger(y)$, then a post-processing network is trained to remove noise and limited-view artefacts, i.e.,
\[
x_\mathrm{rec}=G_\theta (x_0)= G_\theta (A^\dagger(y)).
\]
We will use here the same U-Net architecture for the network $G_\theta$ as in the MC-PD and PD-DEQ models.

Secondly, we will compare our methods to a version of MC-PD that does not use weight-sharing and thus is close to the original learned primal dual (LPD) \cite{Adler2018}, we will also refer to this model as LPD in the following. This network clearly has the most expressive power as each iteration has its own network weights in primal and dual space, and hence we will consider this here as the state-of-the-art to compare our method to. We note, that also here the networks are chosen with the same U-Net architecture. Training of the U-Net takes only 12 minutes and training the LPD takes 120 minutes.

\section{Results and discussion}\label{sec:resultsDisc}

We will evaluate the proposed methods for their quantitative and qualitative performance, but also the ability to provide scalable and robust reconstructions. Followed by a discussion on the DEQ models.

\subsection{Results for 2D simulated data}

We first evaluate the performance on the test data as described in sec. \ref{sec:2Ddata}, that is consistent with the training data, i.e., in a resolution of $80\times 128$ with 1\% noise in the measurements.  For each algorithm we have trained 3 instances and chose the one with the highest validation error to compute the quantitative measures PSNR and SSIM on the test set of $N=93$ samples, see Table \ref{tab:errors} for the obtained values. 

\begin{table}[!h]
\centering
\caption{\label{tab:errors} Quantitative measures (PSNR and SSIM) for all considered methods on the test data ($N=93$).}
\begin{tabular}{r | c | c | c | c | c}
 & U-Net & LPD &  MC-PD &  PD-DEQ & Hybrid  \\ 
\hline
PSNR & 21.95 & 29.06 & \bf{29.34} & 22.1 & 24.27 \\ 
SSIM & 0.943 & \bf{0.991} & \bf{0.991} & 0.959 & 0.974  \\
\end{tabular}
\end{table}

There are two interesting observations that we can make. First of all, LPD and MC-PD perform very similar even though LPD does not use weight-sharing as the MC-PD does. This is largely consistent with the literature, that the weight-shared version performs similarly well as its counterpart. Interestingly, in this instance MC-PD was able to achieve even a slightly higher PSNR, but SSIM values are the same. 
Secondly, we notice that the constrained version PD-DEQ has a significant drop in PSNR. This is an unfortunate result, especially when compared to U-Net we only see a slight improvement. Nevertheless, we can observe a clear improvement in SSIM over post-processing with U-Net. The hybrid approach that combines MC-PD and PD-DEQ is able to improve both quantitative values, but can not reach the performance of the unconstrained version. This is in contrast to previous literature, which largely report that DEQ models can achieve nearly as good (or better) results as their unconstrained versions \cite{gilton2021deep}. We point out here, that in our case we are in a severely ill-posed limited-view setting. Thus, we attribute this drop in performance to the added complications of the limited-view problem in PAT, as we will discuss further below.

Let us now examine the qualitative performance for one sample from the test set shown in Figure \ref{fig:recons_test}. We can immediately notice that the most striking differences are in the recovered quantitative values of each method. None, of the methods manages to recover the contrast very well in the left upper vessel, LPD and MC-PD perform best in this regard. The two constrained DEQ methods primarily show a loss of contrast in the center, whereas U-net has a strong overestimation as well as stark jumps in the contrast. In terms of limited-view artefacts, which are primarily at the bottom corners, all methods perform very well, only U-net and PD-DEQ show some slight blurring in the lower vessels. Generally, the visual appearance is remarkably good given the limited-view setting. In particular with respect to vessel shapes, the other three LPD, MC-PD, and PD-Hybrid perform very similarly and provide overall visually good results.

\subsection{Scalability}
The primary motivation to use approximate models and the DEQ formulation is to enable a scalable learned iterative method, which means applicability to large image sizes and short computation times to enable reasonable training times. Consequently, we will examine next how the different algorithms scale with respect to image size and computation times. 

\begin{figure}
    \centering
    \includegraphics[width=1\linewidth]{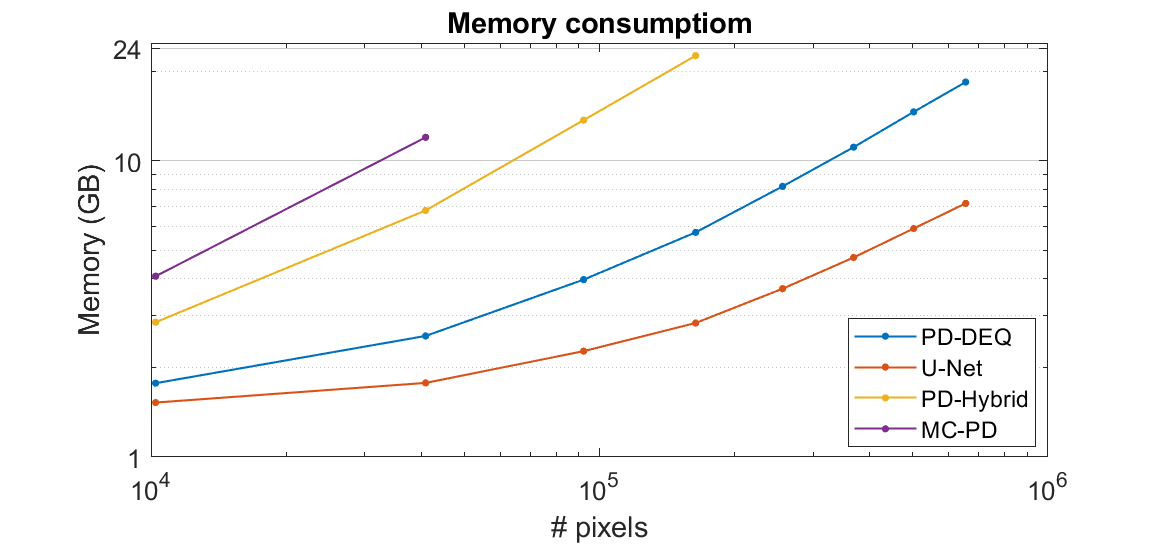} 
    \includegraphics[width=1\linewidth]{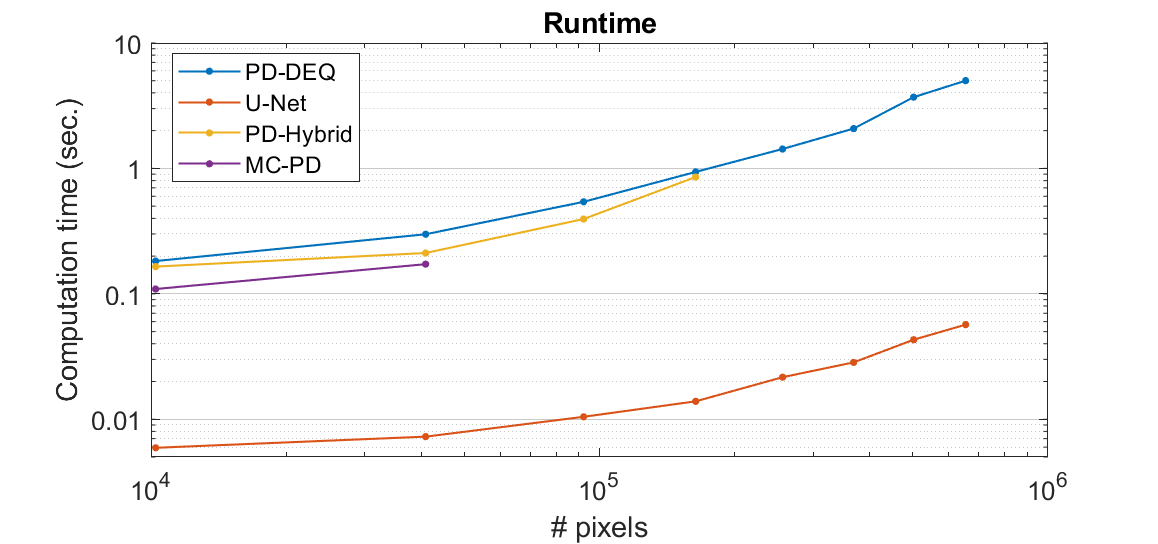} 
    \caption{Scalability tests for considered methods. The top shows memory consumption and the bottom plot shows runtime of the forward pass.}
    \label{fig:scalability}
\end{figure}

For this we have progressively increased the considered image size from the initial $80\times 128 \approx 10^4$ pixels to $640\times 1024 \approx 6\cdot 10^5$ pixels, while keeping the network architectures fixed. We note, that the measurement data for the largest case is consequently of size $1280\times 1024$. In Figure \ref{fig:scalability} we can see how the different methods scale on increasing data size. First of all, we see that the memory consumption of the PD-DEQ model and U-Net scale well with image size and can be applied on the final image size. While MC-PD runs out of memory already at a factor 3 of the initial image size. PD-Hybrid can leverage the reduced consumption and scales slightly better. We have omitted LPD here, as it would behave similar to MC-PD. We note, that better scalability with respect to memory consumption can be easily achieved by considering smaller networks (especially fewer channels) and for MC-PD and PD-Hybrid fewer iteration, wheres the DEQ memory consumption only depends on network size and is iterate independent. 
In terms of runtime, U-Net clearly outperforms all other methods as it only requires one reconstruction with the fast inverse and one network evaluation. The iterative approaches all perform similar, with a slight but not significant overhead for the PD-DEQ model compared to PD-Hybrid and MC-PD. Up to a factor 4, i.e., image size of $320\times 512$, we can achieve a runtime of less than 1 sec. for the PD-DEQ model. Here, a further speed up can be easily achieved by considering fewer unrolled iterations. We note that 1 second for the forward pass translates to roughly 7 hours for the training time with 25000 training iterates. Thus, even the largest image size for PD-DEQ would have still acceptable training times with $\sim$35 hours.

Finally, we examine how the models perform when applied to larger image size than trained on. For this test we used the full image size in the data generation and created samples of size $160\times 512$ for testing. The results are shown in Figure \ref{fig:highRes_eye}. 
We can see that the networks work well when applied to a larger image size than trained on. While all iterative methods provide good image quality consistent with the results on the test set, we can see that U-Net post-processing creates some residual artifacts in the empty areas. Similarly as in the previous case, the constrained methods can not fully recover the contrast in the high absorbing region and some limited-view artifacts can be seen by slight blurred out features in the bottom corners.

{\newcommand{\showpic}[3]{%
\begin{tikzpicture}[spy using outlines={circle, magnification=2, size=2.5cm, connect spies}]%

\draw (0,0) node [anchor=north] {\includegraphics[width=8cm]{recons/8large_#3}};%
\draw (0,0) node [anchor=south, yshift=0pt] {\phantom{f}#1\phantom{g}};%
\draw (0,0) node [anchor=south, yshift=-10pt] {\phantom{f}#2\phantom{g}};
\spy on (3.3cm,-2.15cm) in node at (3.5cm, -3.5cm);
\spy on (0.8cm,-1.5cm) in node at (-1.5cm, -3.5cm);
\end{tikzpicture}
}

\begin{figure*}[t!]
\centering
\showpic{Phantom}{}{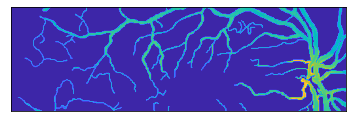}%
\hfill
\showpic{U-Net}{PSNR: 24.64 dB / SSIM: 0.945}{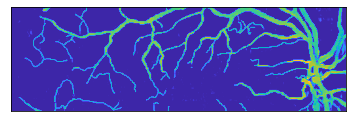}%
\\
\vspace{-0.2cm}
\showpic{MC-PD}{PSNR: 27.75 dB / SSIM: 0.982}{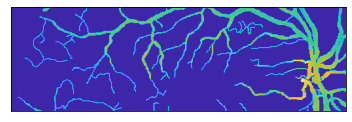}% 
\hfill
\showpic{LPD}{PSNR: 29.61 / SSIM: 0.988}{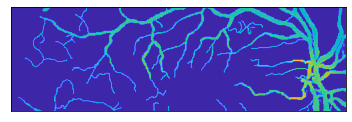}% 
\\
\vspace{-0.2cm}
\showpic{PD-DEQ}{PSNR: 25.28 dB / SSIM: 0.964}{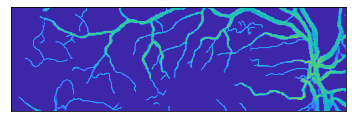}%
\hfill
\showpic{PD-Hybrid}{PSNR: 26.94 / SSIM: 0.977}{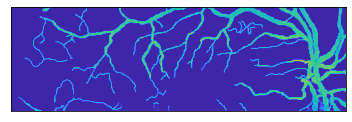}%
\\
\includegraphics[width = 0.8\linewidth]{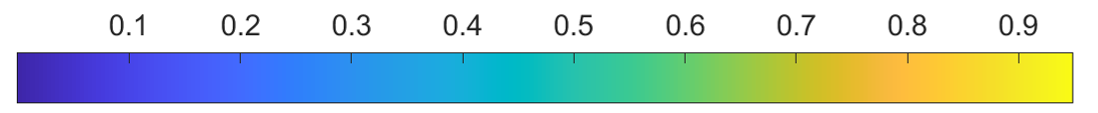}
\caption{\label{fig:highRes_eye} Obtained reconstructions for the considered methods on larger test images of size $160\times 512$, shown with quantitative measures. All images are on the same color scale.  Measurements are taken on the top edge, with 1\% Gaussian noise.}
\end{figure*}}

\subsection{Challenges in the limited-view setting}
In our experiments we have observed that training the DEQ models in the limited-view setting comes with some challenges. First of all, as we can see in the results the DEQ networks do not perform as well as their unconstrained counterparts. This can be in parts attributed to the fact that the constrained networks are limited in their expressivity, similar behavior can be observed for other constrained networks in inverse problems, such as convexity constraints \cite{mukherjee2020learned}. Furthermore, the limited-view setting causes the initial reconstruction to have wrong contrast, due to a loss in energy from only one-sided measurements. To counteract this, we have scaled the input values (by factor 4) to have approximately the correct contrast between in $[0,1]$. Nevertheless, the areas close to the detector are better resolved and have higher contrast than the bottom area. We postulate, that the contractive nature of the DEQ models is not ideally suited to correct in this scenario for this shortcoming and hence we can generally see a loss of contrast in the PD-DEQ and PD-Hybrid reconstructions compared to MC-PD as well as U-Net. We believe this is the main cause for the drop in PSNR. 

\subsubsection{Influence of algorithm parameters}
The theory in Section \ref{sec:convAna} predicts that small values of $\tau,\sigma$ are needed to provide convergence. Indeed, we have observed that with too large $\tau,\sigma>1/(5L)$ the iterations are unstable and training may not be successful. While too small values $<1/(20L)$ will diminish the influence of the updates. Our experiments have shown that between those values good stability of the training and convergence is observed. 

For the DEQ models, the number of iterations plays an important role as well. Here a similar behavior has been observed, too few iterates $<5$ did not provide a sufficient improvement, while too many iterates $>20$ often failed to converge and lead to unstable training. Thus, we have chosen 10 iterates for this study.

\subsubsection{Spectral normalization}\label{sec:spectral}
We have trained the primal update networks $G_\theta$ with and without spectral normalization. While spectral normalization stabilizes the training and larger values of $\tau,\sigma$ were possible to choose, we have also observed that the training got more often stuck in bad local minima during training. Without spectral normalization early iterates may be unstable, but with small enough $\tau,\sigma$ we were able to reliably train the networks and better performance was observed for most training runs. 
As discussed in sec. \ref{sec:DEQ-train}, even without spectral normalization it has been observed that the training implicitly enforces the contractiveness and a stable training is possible. Additionally, spectral normalization adds a computational overhead. Consequently, we have performed the final training and tests without spectral normalization. 

\subsubsection{Training without the model correction}
An ablation study for the DEQ models has been performed by training without the model correction network $F_\phi$, which is closer to the original DEQ model. Unfortunately, we were not able to obtain sufficiently satisfying results. While the instability issues were exacerbated, the networks failed to find good local minima. Smaller values for $\tau,\sigma<$ helped to stabilize the training, but prevented good results. In our experiments we have not been able to exceed a PSNR of 15 dB in the validation set. This suggests that the model correction network in the dual space is indeed necessary in this setting.

\subsection{Convergence of the PD-DEQ model}
\begin{figure}
    \centering
    \includegraphics[width=0.49\linewidth]{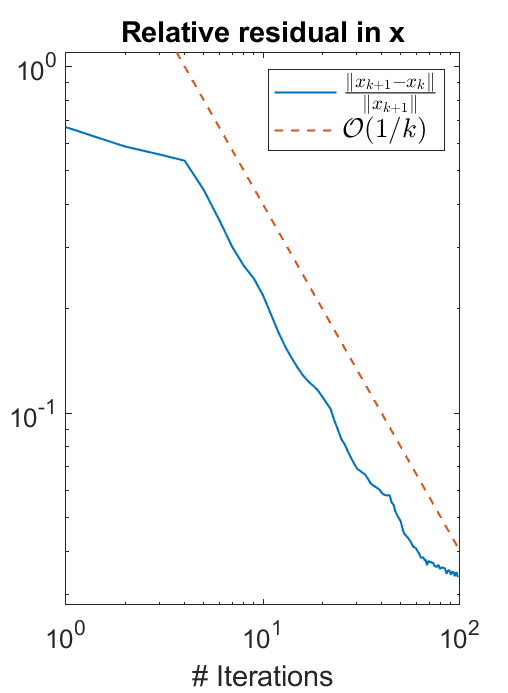} 
    \includegraphics[width=0.49\linewidth ]{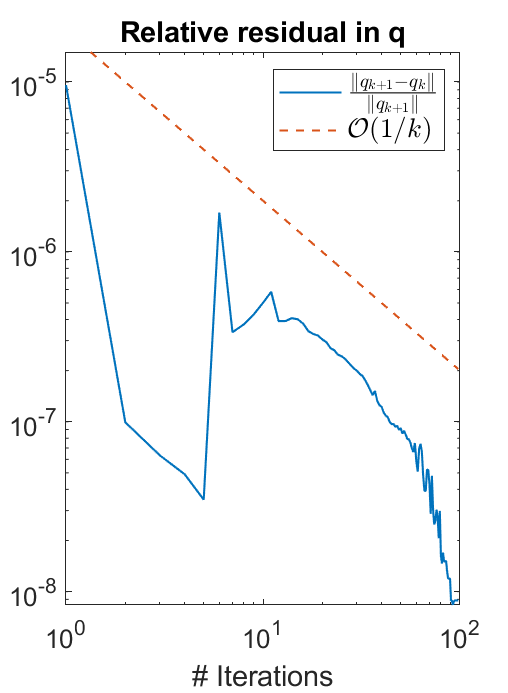} 
    \caption{Convergence of the residuals in the PD-DEQ formulation compared to a rate of $1/k$. Networks are trained for 10 iterates and tested for 100.}
    \label{fig:convergence}
\end{figure}

We have numerically tested how the PD-DEQ model behaves when we continue to run the iterates after the maximum of 10 training iterates has been hit. The theory would predict that the iterates are contractive in the primal variable, while the dual variable is uniformly bounded. This can be observed in Figure \ref{fig:convergence} for the test example shown in Figure \ref{fig:recons_test}. The residual in $x$ does indeed contract with a rate of $1/k$, while the dual does not contract in the beginning, but has considerably smaller relative norm, after the initial 10 iterations also the dual variable contracts with a rate higher than $1/k$ as required for the fixed-point convergence. Nevertheless, the best reconstruction quality is obtained at the initial 10 iterations, while after those a further loss of contrast occurs.

\section{Conclusions}\label{sec:conclusions}
This study aimed to formulate a joint model-correction and learned iterative reconstruction that can be scaled to larger image sizes. We have formulated a model-corrected learned primal dual (MC-PD) and a corresponding deep equilibrium formulation (PD-DEQ). MC-PD provides excellent reconstruction quality, but requires higher memory demand due to the linear growth of the computational graph with number of iterates. PD-DEQ provides excellent scalability, but with quantitatively worse results. Additionally, we have proposed a PD-Hybrid method that balances both, improved reconstruction quality under reduced memory requirements. Further modifications on the hybrid approach are expected to improve quantitative results.

Secondly, this study aimed to provide further methodological and theoretical insights into learned iterative reconstructions and model corrections as well as the use of DEQ models in a limited-view setting. While our PD-DEQ implementation did not perform as well as the unconstrained version we also observed that a DEQ model without the model correction component could not be trained successfully at all. This suggests the importance of including a model correction and provides promising directions for future research.

\bibliographystyle{unsrt}
\bibliography{inv_prob}

\end{document}